\documentclass[5p,,preprint,10pt,twocolumn]{elsarticle}

\usepackage{lineno,hyperref}
\usepackage{dblfloatfix}
\usepackage{graphicx}
\usepackage{multirow} 
\usepackage{booktabs}
\usepackage{amsmath}
\usepackage{amssymb}
\usepackage{mathtools}
\usepackage{booktabs}
\usepackage{caption} 
\usepackage{overpic}
\usepackage{color}
\usepackage[dvipsnames]{xcolor}
\usepackage{float}
\restylefloat{table}
\modulolinenumbers[5]

\journal{Ultramicroscopy}

\newif\ifdraft
\draftfalse
\drafttrue

\newcommand{\comment}[1]{}
\newcommand{\parag}[1]{\vspace{-2mm}\paragraph{#1}}

\ifdraft
 \newcommand{\PF}[1]{{\color{red}{\bf PF: #1}}}
 
 \newcommand{\CH}[1]{{\color{blue}{\bf CH: #1}}}
 
 \newcommand{\OA}[1]{{\color{BlueGreen}{\bf OA: #1}}}
 
 \newcommand{\AM}[1]{{\color{Dandelion}{\bf AM: #1}}}
 
 \newcommand{\EO}[1]{{\color{violet}{\bf EO: #1}}}
 
\else
 \newcommand{\PF}[1]{}
 
 \newcommand{\CH}[1]{}
 
 \newcommand{\OA}[1]{}
 
 \newcommand{\AM}[1]{}
 
 \newcommand{\EO}[1]{}
 
\fi

\newcommand{\mC}[0]{{\cal C}}

\newcommand{\norm}[1]{\left\lVert#1\right\rVert}

\begin{document}

\begin{frontmatter}

\title{3D Reconstruction of Curvilinear Structures with Stereo Matching Deep Convolutional Neural Networks
}

\author[cvlab]{Okan Alting\"ovde}
\author[cvlab,lsme]{Anastasiia Mishchuk}
\author[lsme]{Gulnaz Ganeeva}
\author[cime]{Emad Oveisi}
\author[lsme]{Cecile Hebert}
\author[cvlab]{Pascal Fua}

\address[cvlab]{Computer Vision Laboratory, École Polytechnique Fédérale de Lausanne (EPFL), Switzerland}
\address[lsme]{Electron Spectrometry and Microscopy Laboratory, École Polytechnique Fédérale de Lausanne (EPFL), Switzerland}
\address[cime]{Interdisciplinary Centre for Electron Microscopy, École Polytechnique Fédérale de Lausanne (EPFL), Switzerland}


\begin{abstract}

Curvilinear structures frequently appear in microscopy imaging as the object of interest. Crystallographic defects, i.e dislocations, are one of the curvilinear structures that have been repeatedly investigated under transmission electron microscopy (TEM) and their 3D structural information is of great importance for understanding the properties of materials. 3D information of dislocations is often obtained by tomography which is a cumbersome process since it is required to acquire many images with different tilt angles and similar imaging conditions. Although, alternative stereoscopy methods lower the number of required images to two, they still require human intervention and shape priors for accurate 3D estimation. We propose a fully automated pipeline for both detection and matching of curvilinear structures in stereo pairs by utilizing deep convolutional neural networks (CNNs) without making any prior assumption on 3D shapes. In this work, we mainly focus on 3D reconstruction of dislocations from stereo pairs of TEM images.

\end{abstract}



\end{frontmatter}

\section{Introduction}

Transmission Electron Microscopy delivers a 2D projection of the sample under consideration while many areas of biological and physical sciences now  require 3D reconstructions of nano-sized objects. In particular, curvilinear objects such as dislocations~\cite{Amelinckx74,Hirsch2006a} are of great interest in material science. 

The traditional approach to recovering the third dimension is to tilt the sample with respect to the electron beam, to capture dozens of views, and to perform tomographic reconstruction~\cite{DeRosier1968,Weyland15,Midgley2009,Sharp2008,Barnard2006a}. This is time consuming and usually delivers a data volume from which the dislocations still have to be traced manually. An alternative is to use stereography instead. It only requires a few images.  Unfortunately, state-of-the-art methods~\cite{Jacome12,Jacome18,Oveisi17a,Oveisi18a} still involve substantial manual intervention for line detection and matching. 


\begin{figure*}[!t]
\centering
\begin{tabular}{c}
\begin{overpic}[width=\textwidth]{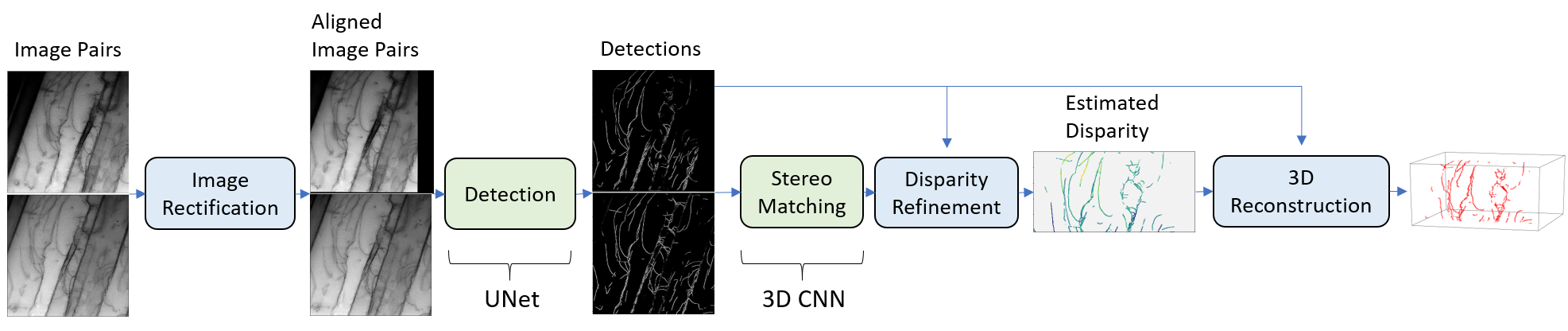}
\end{overpic}
\end{tabular}
\vspace{-4mm}
\caption{Stereo reconstruction processing steps. In the first step, images are aligned such that the tilt axis is approximately vertical and at the image center. Dislocation segments are detected using UNet \cite{Ronneberger15} in the second stage. Later, features of UNet are provided to a 3D CNN in order to compute horizontal disparity between corresponding dislocation segments. At the final stage, depth value for every detected dislocation pixel is computed using disparity estimations. Green blocks in the depicted figure represent the stages that are learned from annotated training data.}
\label{fig:pipeline}
\end{figure*}

In this paper, we present a fully automated stereo-based approach that only requires two images and directly outputs the 3D reconstructions in usable form. Its accuracy nevertheless is similar to that of tomography. Fig.~\ref{fig:pipeline} depicts our proposed pipeline. Its key component is a deep network that jointly detects dislocations in individual images and matches them across images to recover their 3D structure,  while being end-to-end trainable in a semi-supervised way. The latter is important because ground truth is rarely, if ever, available. A key improvement over earlier methods is that the detection and matching steps are performed jointly instead of sequentially. In addition to eliminating manual intervention, it stops the algorithm from finding spurious dislocations.

In short, our pipeline delivers reconstruction accuracy comparable to that of tomography while reducing the imaging and processing times. Moreover, it automatically detects dislocations 
and saves us from filtering the final reconstruction with manually set thresholds.

\section{Method}

Depth information can be recovered from two or more images by triangulating the 2D projections of specific 3D points that have been matched across images. Such points are known as feature-points. They can either be sparsely sampled across the images or form linear structures. 

In the early days of computer vision, computational power was very limited and such feature-based stereo-reconstruction techniques arose from the need to limit the computational requirements by operating only on a small fraction of the image. An additional strength of these algorithms was that, because features could be detected to subpixel-accuracy, good reconstruction accuracy could be achieved. For example, as early as in 1988, the system of~\cite{Hannah88} achieved precisions in the order of half a pixel in disparity, which has not been improved much since~\cite{Schonberger17}. Unfortunately, it only matched a very small proportion, typically less than 1\%, of the image points and in relatively narrow baseline stereo pairs. Among these approaches, contour-based ones became popular in the 1980s~\cite{Medioni85,Ayache87,Meygret90} because they made it possible to reason about contour continuity and impose additional geometric constraints. \comment{These algorithms handled mostly line segments and were later extended to operate on curves whose projections might be influenced by the viewing angle~\cite{Li03a}.} However, with the advent of algorithms based on correlation~\cite{Fua93a}, graph-cut~\cite{Boykov01b}, or deep-learning~\cite{Yang18d,Ji18,Huang18b,Kar17} that could match much larger fractions of the images at a computational cost that modern hardware can easily handle, all these feature-based techniques have fallen out of favor and there has been remarkably few new developments in the last 20 years. 

\begin{figure*}[!t]
\centering
\begin{overpic}[width=1\hsize]{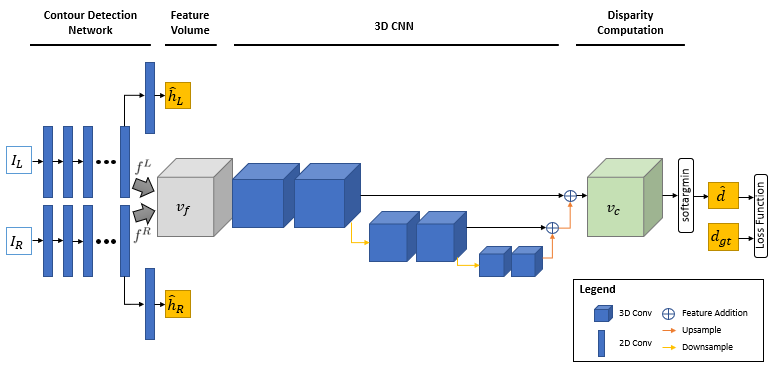}
\end{overpic}
\caption{Stereo Contour Detection and Matching Network architecture. Layers with learnable parameters are shown in blue. Feature volume, $V_{f}$, and cost volume, $V_{c}$, are depicted as gray and green 3D volumes respectively.}
\label{fig:network_architecture}
\end{figure*}

Conceptually, our stereo reconstruction pipeline follows the classical paradigm and comprises the four following steps: Image rectification, detection of the linear structures, stereo matching, and finally triangulation to turn the 2D contours into 3D dislocation models. These fours steps are depicted by Fig.~\ref{fig:pipeline}. In the earlier era described above, they would have been performed sequentially and independently from each other. However, with the advent of Deep Learning~\cite{LeCun15}, it has become both possible and desirable to merge detection and matching and to have them be performed jointly by the end-to-end trainable deep net, such as the one shown in Fig.~\ref{fig:network_architecture}.

In the remainder of this section, we first discuss our approach to detection and matching of dislocations in 2D images in the ideal case where these images are properly aligned. We then discuss how they can be {\it rectified} so that they become aligned and finally how to generate fully 3D dislocations from the 2D projections. 

\subsection{Detecting and Matching Dislocations}
\label{sec:pipeline}

In two-view stereo, given a point in the first image, recovering its 3D position involves finding the corresponding point along the so-called {\it epipolar} line. If the two images are coplanar, that is, they were taken such that the second camera is only offset horizontally compared to the first one, then each pixel's epipolar line is horizontal and at the same vertical position as that pixel. This also applies to pairs of TEM images because they are orthographic projections of the same object on a fixed detector/camera acquired while the object is being tilted in the microscope.

In this section, we will assume this to be the case, that is, for a point, $p=(x,y)$ in the first image,  the corresponding point $q$ is at location $(x+d, y)$ in the second image, where $d$ is the {\it disparity} and proportional to the distance to the image plane. We will show in the following section how the images can be rectified to make this be true. 

\subsubsection{Network Architecture}
\label{sec:architecture}

To detect the dislocations in the 2D images and compute disparities for each one of their points, we use the deep network depicted by Fig.~\ref{fig:network_architecture}. Its first stage consists of a 2D \emph{contour detection network} $\mC$ in a Siamese configuration \cite{Bromley93}. The feature maps produced by $\mC$ for both views are combined into the \emph{feature volume}, which is then refined by a 3D CNN into a \emph{cost volume}.

We describe each building block of architecture in more detail below. 


\begin{figure}[!b]
\centering
\begin{overpic}[width=0.90\hsize]{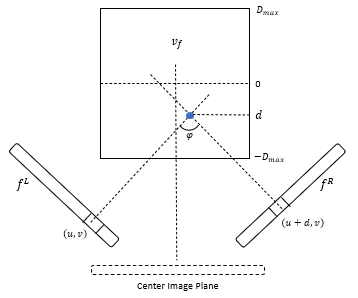}
\end{overpic}
\vspace{0.5cm}
\caption{Feature volume construction is shown from top view. For each pixel in left image, features of this pixel and candidate pixels for a match in the right image are back-projected to 3D and concatenated to form feature volume. For a specific pixel at location $(u, v)$ in left image and a candidate pixel at location $(u + d, v)$, back-projected point in feature volume is depicted as blue circle in the figure. The value $d$ for the given pixel is the disparity level that defines the depth of the pixel with respect to the center image plane. }
\label{fig:feature_volume}
\end{figure}

\begin{figure*}[!t]
\centering

\begin{overpic}[width=0.85\hsize]{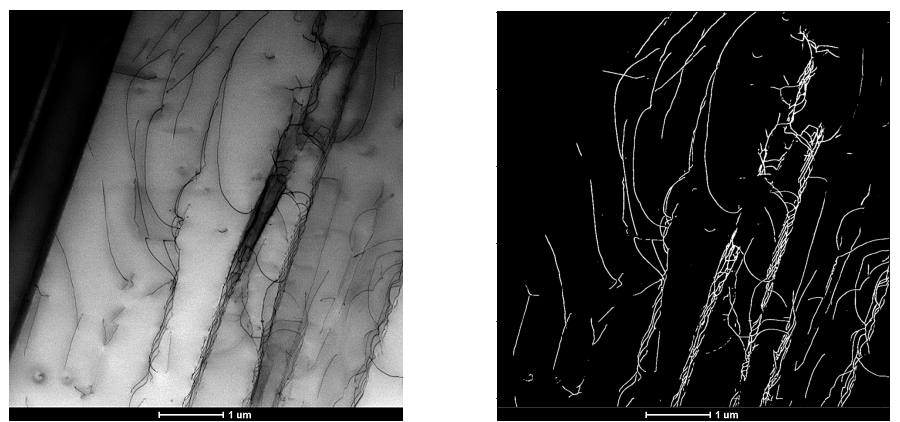}
\end{overpic}\\
(a) \hspace{8cm} (b)
\caption{Representative 2D ABF STEM image of dislocations in a TiAl super-alloy thin foil (a) and corresponding detected dislocations (b). It might be seen that detection network manages to detect dislocations having complex shapes in noisy images while still being able to reject unrelated dark sections of the image.}
\label{fig:detection}
\end{figure*}

\parag{Pre-trained contour detection network} 
\label{sec:detectNet}

 $\mC$ extracts contour-specific features. 
 We denote the feature maps by $f^L$ and $f^R$ and contour detections by $\hat{h}^{L}$ and $\hat{h}^{R}$ for the left and right images, $I_L$ and $I_R$, respectively. $\mC$ is trained with binary cross-entropy loss. In Fig. 4, we show an example detection output of $\mC$.

\parag{Feature Volume}
As in many existing architectures \cite{Kendall17a}, we construct a feature volume $V_{f}$ shown in Fig. 3, by replicating the left and right feature maps once for every possible disparity value, and concatenating them into a tensor of size $2 \cdot \textit{feature size} \times \textit{height} \times \textit{width} \times \textit{disparity range}$, so that $V_{f}(\cdot,u,v,d)$ is a concatenation of features $f^L(\cdot,u,v)$ and $f^R(\cdot,u+d,v)$, where $u$ and $v$ are coordinates of the left image and $d$ is disparity. This enables the evaluation of match hypotheses $(u,v,d)$ by means of a 3D CNN as follows. 

\parag{3D CNN for Matching Cost Estimation} 
\label{sec:disptNet}

A 3D CNN is used to compute a cost volume $V_c$, shown in green in Fig.~\ref{fig:network_architecture}, from the feature volume $V_f$. $V_c$ is of  size  $\textit{height} \times \textit{width} \times \textit{disparity range}$ and an entry $V_c(u,v,d)$ encodes the cost of matching a feature $f^L$ at position $u,v$ in the left feature map, to a feature $f^R(u + d, v)$ of the right feature map.
3D CNN consists of multiple stacked hourglass modules composed of 3D $3 \times 3  \times  3$ convolution layers, shown in blue volumes in the center of Fig.~\ref{fig:network_architecture}.

\paragraph{Disparity Computation}  

Optimum disparity for a pixel at location $(u, v)$ is the one with lowest matching cost which is encoded in cost volume, $V_{c}$. Therefore, a disparity estimate is the index of the minimum element along disparity axis of $V_{c}$ for each pixel, which could be found using the \textit{argmin} operation. However, \textit{argmin} is not differentiable and can therefore not be used as part of neural network architecture.Instead we use the differentiable \textit{soft-argmin} defined as
\begin{align}
&\hat{d}_{uv} = \sum^{d_{max}}_{d = 0}{d \cdot \sigma(-V_{c}(u,v,d))}
\quad 
\label{eq:softargmin}
\end{align}
where $\sigma(.)$ is the $softmax$ operation, $d$ is disparity level and  $V_{c}(u,v,d)$ is matching cost of pixel at location $(u,v)$ for disparity $d$ to be able to train our network. 

\subsubsection{Loss Function} 
\label{sec:losses}

We train our stereo matching network with a loss function that is a weighted sum of three loss terms
\begin{align} \label{eq:compLoss}
L = \gamma_{disp} L_{disp} +\gamma_{var} L_{var} +\gamma_{warp} L_{warp} \; ,
\end{align} \\
where $L_{disp}$, $L_{var}$, and  $L_{warp}$ represent a disparity , variance, and warp loss, which we describe in more detail below.  

\subparagraph{Disparity Loss.}

When the ground-truth disparity , $d_{gt}$, is available, we penalize inconsistencies in the prediction $\hat{d}$ by means of the smooth L1 loss 
\begin{align} \label{eq:dispLoss}
L_{disp} = \frac{1}{|\mathcal{A}|}\sum_{a \in \mathcal{A}}{
	\begin{cases}
	0.5 \cdot {e_d}^2, & \text{if } e_d < 1\\
	e_d  - 0.5, & \text{otherwise} 
	\end{cases}}
	\; ,
\end{align}
where $e_d = \norm{\hat{d}- d_{gt}}$  and $\mathcal{A}$ is the set of pixels belonging to dislocations with available disparity annotations. 


\begin{figure}[!b]
\centering
\begin{overpic}[width=0.85\hsize]{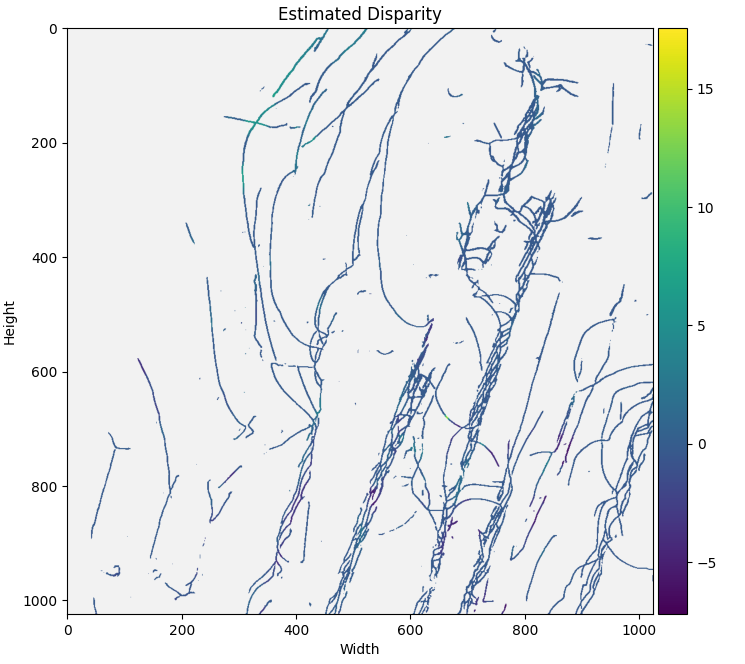}
\end{overpic}
\vspace{0.5cm}
\caption{Estimated disparity of detected dislocations shown in Fig. 4. The disparity map is made using pair of stereo images with 4 degrees of tilt between two images. Depth axis is towards image plane. Dislocations segments with bright colors are closer to the image plane. Width, height and disparity values are all in pixels. Best viewed in color.}
\label{fig:dislocations_disparity}
\end{figure}

\subparagraph{Variance Loss.}

While minimizing $L_{disp}$ encourages the network to find the correct disparity value, it is also important that it delivers a low variance around that ground truth value so that the estimates are accurate. Therefore, we define variance loss
\begin{align}\label{eq:varLoss} 
L_{var} = \frac{1}{|\mathcal{A}|}\sum_{a \in \mathcal{A}}{\hat{v}} \; ,
\end{align} 
where  $\quad \hat{v}_{uv} = \sum^{d_{max}}_{d = 0}{(d - d_{gt})^2 \cdot \sigma(-V_{c}(u,v,d))}$. \\

Disparity estimator defined in Eq. \ref{eq:softargmin} may be seen as probability-average of possible disparity levels by taking $\sigma(-V_{c}(u,v,d))$ term as probability of disparity level $d$. In this case, $\hat{v}$ becomes expectation of the squared deviations of disparity around the disparity ground truth. Minimizing $L_{var}$ therefore minimizes the variance of the estimate $\hat{v}$ around $d_{gt}$.

\subparagraph{Warp Loss.}

Given the detection map of curvilinear structures in the right image, $\hat{h}^R$ and the disparity, $\hat{d}$, the left detection map, $\hat{h}^L$ can be reconstructed by shifting pixels with their corresponding disparity values. This operation is called \textit{warping} and we incorporate it in our warp loss
\begin{align}\label{eq:warpLoss} 
L_{warp} &=  L_{bce}(warp(\hat{h}^{R}, \hat{d}), \enspace h_{gt}^{L}) \; . \;
\end{align}
where $L_{bce}$ is the binary cross entropy loss. Minimizing $L_{warp}$ enforces consistency between $\hat{d}$, $\hat{h}^R$ and $\hat{h}^L$.  \\


\begin{figure*}[!t]
\centering
\begin{overpic}[width=0.8\hsize]{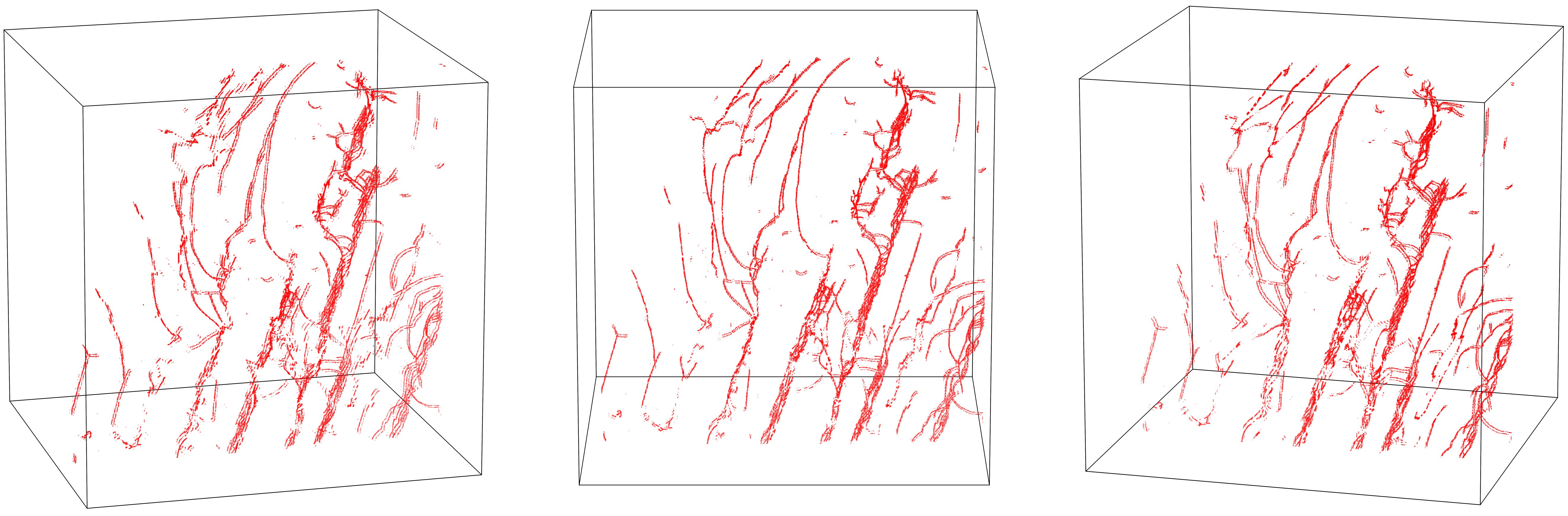}
\end{overpic}
\vspace{0.5cm}
\caption{3D Reconstruction of dislocations shown in Fig. 4.}
\label{fig:dislocations_3d}
\end{figure*}

Overall performance of 3D CNN depends on its capability to localize minimum cost disparity level in $V_{c}$ for each pixel. In other words, it is required to have low variance probability distributions over estimated disparity values to confidently make predictions. In this sense, complementing disparity loss with variance and warp loss terms allows us to further constrain network training to have cost estimations not only centered at ground truth level but also having lower variances around it. This is especially important when training data is not abundant and cumbersome to annotate which is the case for electron microscopy imaging. In result section, we present an ablation study on components of loss function to show strong evidence in favor of using complemented loss function over naive disparity loss.  

\subsection{Image Rectification}

In practice, the images are never perfectly aligned.  Although the viewing angle can be controlled up to a degree, the region of interest is not necessarily located around the rotation axis and need to be centered manually during image acquisition. Manual search of the region causes translational misalignment between consecutive images acquired. Moreover, it is not always possible to maintain exact vertical tilt-axis due to a non-perfectly compensated image rotation which may yield deviations up to 2$^{\circ}$ from the vertical axis. Thus, we write the relationship between image coordinates of corresponding points $<p, q>$ in the original unaligned images as  
\begin{align}
(x_{p}, y_{p}) &= (x_{q} + d_{q}\cos{\theta} - t_{x}, y_{q} + d_{q}\sin{\theta} - t_{y}) \; , \nonumber \\
(\Delta x_{pq}, \Delta y_{pq}) &= (d_{q}\cos{\theta} - t_{x}, d_{q}\sin{\theta} - t_{y}) \; , \label{eq:alignment_offsets}
\end{align}
where $\theta$ is the angle the tilt-axis makes with the vertical axis on the image plane, $d_{q}$ is disparity of point $q$, $(t_{x}, t_{y})$ are translational shifts that occurred during acquisition on axes x and y of image plane respectively and $(\Delta x_{pq}, \Delta y_{pq})$ are total displacements on image plane between correspondences prior to alignment. 

Hence the image transformation that corrects this misalignment is  
\begin{align}
T &= [I \ | \ t] \nonumber \\
(\hat{x_q}, \hat{y_q})^{T} &= T_{[2 \ \times \ 3]} (x_q, y_q, 1)^{T}
\label{eq:alignment_proj}
\end{align}
where $I$ is the $2 \times 2$ identity matrix and $T$ is the $2 \times 3$ transformation matrix that translates point coordinates expressed in projective coordinates. 

We automatically detect and match Scale Invariant Feature Transform (SIFT)~\cite{Lowe04} keypoints to construct sparse correspondences between left and right images. These points are used to estimate the transformation $T$ by iteratively solving a linear system of equations. At each iteration, outliers are eliminated using random sample consensus (RANSAC) \cite{Bolles86}. After applying $T$ to the right image, the total displacement between correspondences, $(\Delta x_{p\hat{q}}, \Delta y_{p\hat{q}})$ becomes $(d_{q}cos{\theta}, d_{q}sin{\theta})$ and $\theta$ can be computed as
\begin{align}
\theta = \arctan(\frac{\Delta y_{p\hat{q}}}{\Delta x_{p\hat{q}}}) \; .
\label{eq:alignment_theta}
\end{align}

Finally, the images are aligned by applying the rotation matrix  
\begin{equation}
R=
\begin{pmatrix*}[l]
    \cos{\theta} & -\sin{\theta}  \\
    \sin{\theta} & \phantom{-}\cos{\theta}  \\
\end{pmatrix*}
\end{equation}
to the transformed right image. Once this is done, the epipolar lines are parallel to the horizontal axis of the images, which is a prerequisite for inputs to our deep network as discussed in the beginning of this section. 


\begin{figure*}[!t]
\centering
\begin{overpic}[width=0.98\hsize]{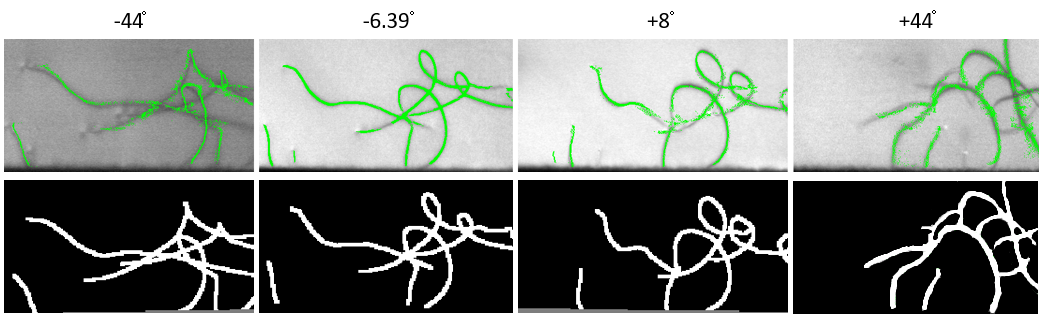}
\end{overpic}
\caption{Re-projections to other views. Estimated 3D dislocations in TiAl sample that are reconstructed using a stereo pair with 8 degree separation are re-projected to views with -44, -6.39, +8 and +44 degrees tilt angles. In the top row of images, four views are shown with green marks being 3D dislocation points re-projected onto raw images. In the bottom row, dislocation ground truth for these views are shown. For all presented views, it can be seen that re-projections of the estimated 3D structure are consistent with raw images and human annotated dislocations ground truth.}
\label{fig:projections}
\end{figure*}

\subsection{Disparity Refinement}
\label{sec:refine}

Disparity values are computed from the estimated cost volume using a soft-argmin operation as shown in Eq.~\ref{eq:softargmin}. In practice, this yields disparities that are within one or two pixels of what they should be. To improve accuracy, we use a standard technique to refine them~\cite{Liang18, Pang17}: We use them to warp the dislocations found in the left image into estimates in the right image, which we then match to the closest dislocation actually found in the right image to refine the disparity value. Warped dislocations are matched pixel-wise to the closest candidate within a 3 pixel distance, which delivers the pixel accuracy required for accurate reconstruction. Dislocation pixels without any suitable candidate in this range are marked as erroneous and removed from the final 3D reconstruction. 

In the result section, we will compare reconstructions obtained with and without this refinement stage and show that, on average, it improves precision.

\subsection{3D Reconstruction}

Having disparity values computed, the 3D shapes of curvilines can be reconstructed by applying triangulation with given viewing angles and the orthographic camera assumption. 

In our stereo setup of verged orthographic cameras which is shown in Fig. \ref{fig:feature_volume}, we define the depth axis same as disparity axis which is parallel to surface normal of the imaginary center image. Given this definition of depth, its relation to the disparity $d$ is
\begin{equation}
\text{depth} = \frac{d}{2sin(\frac{\phi}{2})} \; ,
\label{eq:depth}
\end{equation}
where $\phi$ is the angle between two views. The term $sin(\frac{\phi}{2})$ in Eq. \ref{eq:depth} acts as a scaling factor between depth and disparity. In Fig. \ref{fig:dislocations_3d}, a 3D reconstruction of dislocations from TEM images with 4 degrees between images is shown.


\section{Experiments}

\subsection{Datasets}

For our experiments, we created three stereo dislocation dataset each composed of stereo pairs taken from image sequences of three different materials, representing large diversity of dislocation shapes and contrasts. 
Bright-field images of dislocations were aquired from TiAl alloy, GaN, and high-entropy (Cantor) alloy using transmission electron microscope in scanning mode (STEM). Details of imaging conditions and specimen preparation can be found in Appendix. 

Our dislocation dataset consists of 90 labeled stereo pairs grayscale images with 512x512 pixels resolution split into train, validation and two test subsets. Segmentation masks and disparity maps were manually labelled for the task. The dataset will be publicly available at \emph{www.epfl.ch/labs/cvlab/data/data-dislocations}.


\begin{figure*}[!t]
\centering
\begin{overpic}[width=0.80\hsize]{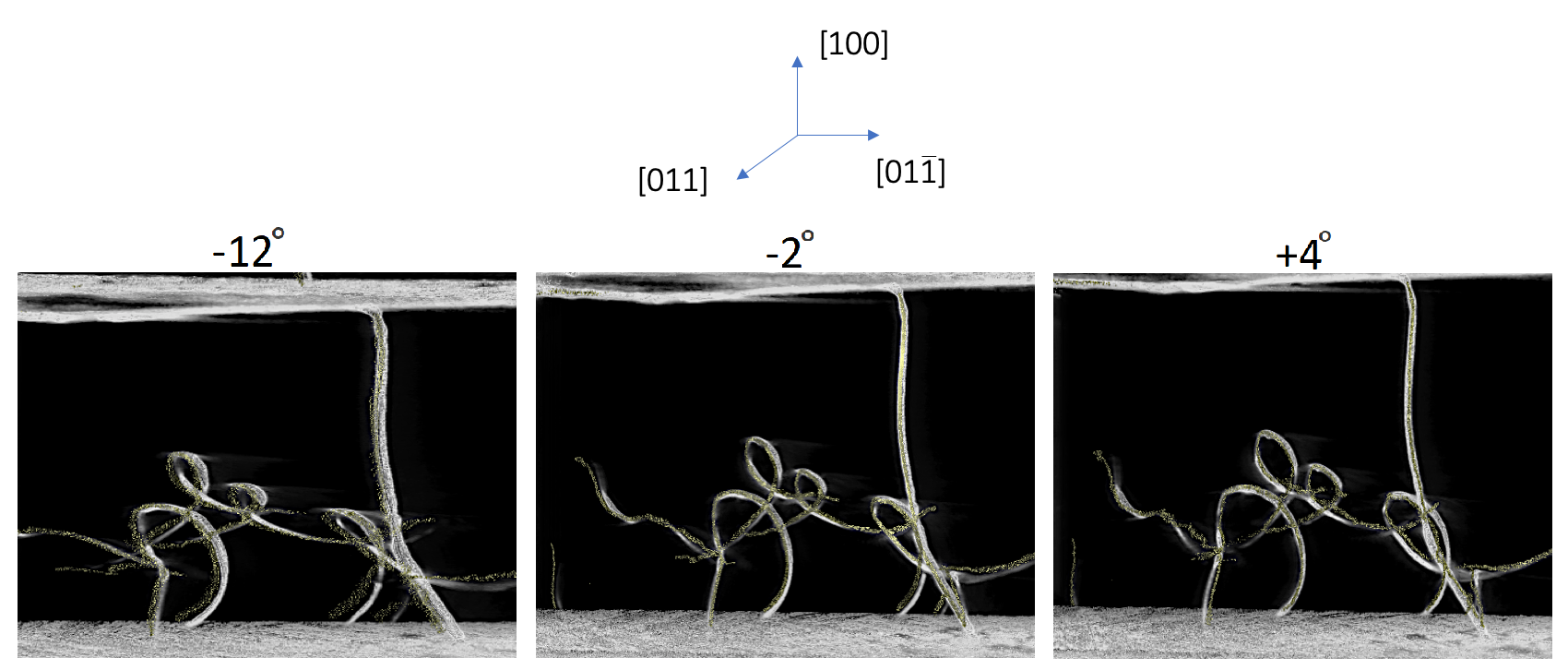}
\end{overpic}
\vspace{0.5cm}
\caption{Comparison of reconstruction results of our approach and tomography. TiAl sample region previously shown in Fig. 7 is reconstructed by tomography (shown in white) utilizing full image sequence with 45 images covering the angular range (-48$^{\circ}$, 50$^{\circ}$) and by our method with stereo images having 2 degrees in between (shown in yellow).}
\label{fig:tomo_comp}
\end{figure*}

\subsection{Results}
When dealing with dislocations in real-world samples, we do not have access to reliable ground-truth because even human annotators experience difficulty reconstructing precise 3D curvilinear structures. We therefore propose two different qualitative ways to evaluate our results and demonstrate their accuracy. We also present an ablation study for different stereo setups with varying tilt angles and loss functions to quantitatively show their effect on the overall reconstruction performance.

\subsubsection{Re-Projection Errors}

Once the 3D dislocations have been reconstructed, we can project them into views that were {\it not} used to perform the reconstruction. When these projections superpose well with the actual dislocations, it is evidence that the 3D reconstructions were correct. In Fig. 7, it is shown for the case where stereo pairs have 8 degrees between them.

\subsubsection{Comparison to Tomography} 

To further evaluate the results Fig. 8 shows qualitative comparison of reconstructed 3D dislocations from  tomography baseline method and our neural network reconstruction. Tomogram was obtained with Inspect 3D Software using Simultaneous Iterations Reconstruction Technique (SIRT). 3D visualisation of the tomogram was performed in Chimera software. In comparison with tomography method, which uses sequence of 45 images covering the angular range (-48$^{\circ}$, 50$^{\circ}$), we are using only one stereo pair with 2 degrees difference between the views. 

Established tomography techniques are mainly brute-force multi-view reconstruction algorithms and rely on geometric constraints to reconstruct 3D shapes. On the other hand, our stereo approach incorporates visual similarities of dislocations segments to lower the ambiguities that may occur when sufficient number of images for reliable reconstruction is not available or cumbersome to acquire. 

To this end, in order to evaluate our method, we compare our stereo approach to two commonly used tomography techniques, Weighted Back Projection (WBP) and SIRT for varying number of images given as input. 
In Fig. 9, it is shown that our stereo reconstruction network outperforms both tomography techniques in low data margin i.e when only a few images are available. To our knowledge, our method is the only fully-automated stereo reconstruction method for curvilinear structures.  We further discuss advantages and drawbacks of the proposed method in discussion section. 

\begin{figure}[!b]
\centering
\begin{overpic}[width=0.77\hsize]{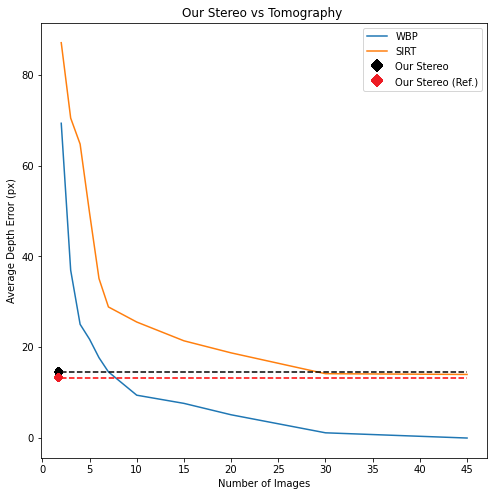}
\end{overpic}
\caption{Two 3D tomography reconstruction techniques are compared to our stereo method across increasing number of images given as input. Images are sampled regularly for each case from the sequence with 45 images spanning angular range (-48$^{\circ}$, 50$^{\circ}$).}
\label{fig:tomo_stereo}
\end{figure}

\subsubsection{Ablation Study}

Human generated annotations are used as groundtruth in order to compute depth and disparity errors of different configurations. We report End-Point-Error (EPE) - the average Euclidean distance between estimated and groundtruth points and the percentage of curviline pixels with disparity error greater than 1, 3 and 5 pixels. 


\setlength{\tabcolsep}{5pt}
\begin{table}[H]
\center
\begin{tabular}{l|c|c|c|c|c|c}
\toprule
& 
2$^{\circ}$&%
4$^{\circ}$&%
6$^{\circ}$&%
8$^{\circ}$&%
10$^{\circ}$&%
12$^{\circ}$%
\\
\midrule
Ours (Raw) & 
18.3&%
21.5&%
15.0&%
14.6&%
22.2&%
28.4%
\\
\midrule
Ours (Refined)& 
18.0&%
20.1&%
14.7&%
\bf{13.2}&%
21.8&%
29.9%
\\
\bottomrule
\end{tabular}
\caption{Depth estimation errors (in pixels) for stereo setups with different angle separations. For all experiments, we trained the network using the loss function that combines all three loss terms. We present results without the refinement stage of Sec.~\ref{sec:refine} and with it.}
\label{tb:results_geo}
\end{table}

To investigate the effect of stereo setup on reconstruction accuracy, stereo pairs that have 2, 4, 6, 8, 10 and 12 degrees between images are used in reconstruction process and corresponding reconstruction errors are shown in Table 1. Reconstruction with small angles results in high depth errors due to heavy discretization although they have low matching error in pair. On the other hand, when the angle between images is large, matching task becomes harder and its error start dominating the total depth error. In our experiments, we found that reconstruction with images having 8 degrees in between addresses this trade-off most efficiently. Moreover, we observe an increase on the measured depth error when angular separation between images increased from 2 degrees to 4 degrees. We attribute it to heavily discretized depth values in reconstructions with extreme narrow angles as 2 degrees. In this regime, depth is encoded into a narrow pixel range on images which renders the matching task effortless since large depth changes mapped to small horizontal shifts. After passing this regime, error increase in disparity estimation can be compensated by performance gains resulted from geometry.

We trained our stereo matching network using different combinations of the loss terms introduced in the method section to tease out their respective contributions. Table 2 shows EPE and percentage of dislocations pixels with errors higher than 1, 3, 5 pixels for different loss term combinations. It may be seen that network trained with unsupervised warp loss alone performs poorly on localization of matching points. While network may be successfully trained with supervised disparity loss, its accuracy increases notably when combined with novel variance loss introduced in this work. We obtained best results when all three loss components combined.


\begin{table}[h]
\centering
\resizebox{\columnwidth}{!}{

 \begin{tabular}{lcccc}
 \toprule
 {Loss function} &  { 1px \% } & {3px \% } & {5px \%} & {EPE px}  \\ 
    \midrule
    {GaN} &  {} & {} & {} & {}  \\ 
    \midrule
    \textbf{Disp+Warp+Var}  & \textbf{40.1} & \textbf{8.3} & \textbf{3.5} & \textbf{1.112}  \\
    \textbf{Disp+Var}  & 40.3  & 8.5 &  \textbf{3.5}  & 1.120   \\
    Disp+Warp  &  41 & 8.6 & 3.8  & 1.238  \\
    Disp  & 52.1  & 9.4 & 4.4  & 1.852      \\ 
    Warp  & 81.6 & 55.3 & 30.2 & 5.034      \\ 

    \midrule
    {TiAl} &  {} & {} & {} & {}  \\ 
    \midrule
    \textbf{Disp+Warp+Var}  & 41 & \textbf{9.6} & \textbf{4.1} & \textbf{1.411}  \\
    Disp+Var  & \textbf{40.7}  & 9.7 & 4.2  & 1.418   \\
    Disp+Warp  & 41.6  & 9.7 & 4.2  & 1.415  \\
    Disp  & 54.9  & 14.2 & 5.8  &   1.755    \\ 
    Warp  & 84.8 & 57.9 & 38.3 &   5.879    \\ 
    
    \midrule
    {Cantor Alloy} &  {} & {} & {} & {}  \\ 
    \midrule
    \textbf{Disp+Warp+Var}  & \textbf{38.2} & \textbf{8.0} & \textbf{3.1} & \textbf{0.982}  \\
    Disp+Var  & 40.5  & 8.4 & 3.3  & 1.101   \\
    Disp+Warp  & 41.3  & 8.7 & 3.6  & 1.322  \\
    Disp  & 49.8  & 9.0 & 4.3  &   1.523    \\ 
    Warp  & 80.4 & 52.4 & 28.5 &   5.011   \\ 

\bottomrule

\end{tabular}
}
\caption{ Comparison of disparity estimation with different combinations of loss terms. Results are presented for three subsets of test data. Stereo pairs with 8$^{\circ}$ angular range is used for the experiments.} 
\label{tb:results_dislocations}

\end{table}


\section{Discussion}

Throughout the last decade, deep neural networks achieved state-of-the-art results in all vision benchmarks for segmentation, object detection, depth estimation and reconstruction. Success of deep networks against its predecessors mainly lies on its capability to learn complex operations solely from available data. Therefore, the performance is dependent to quality, size and diversity of the training data. Immediate challenge in deep network training is to access large datasets. It may be seen from the Fig. \ref{fig:appendix_imp} in appendix section that there are cases even one additional training sample could significantly increase detection accuracy of UNet on test sample. To this end, we constructed largest available annotated dislocation dataset for detection and stereo matching tasks. 

As shown by our experiments, stereo reconstruction of curvilinear structures can be automatized while acquiring accuracy comparable to its established multi-view alternative, tomography. One of the advantages of our approach is that learning-based detection pipeline eliminates the need of manual thresholding of final reconstruction results unlike tomography, while it also enables us to apply structure-aware matching. It is evident that stereo matching of curvilinear structures may be learned from human labeled data by deep convolutional neural networks leveraging visual similarities between structures in stereo. Although relying on visual appearances yields accurate matches, it also limits the tilt range of operation since visual similarity of curvilinear structures rapidly degrades with increasing tilt angle. In our experiments, we have shown that a good performance may be acquired with stereo pairs having 8 degrees in between.

Higher precision reconstructions of curvilinear structures may be obtained via multiview approaches, however, it is crucial to keep required number of images in feasible limits in order to facilitate applicability. Therefore, our research will focus on multiview extention of our joint detection and matching approach in future. 

In conclusion, stereo-vision remains to be an acceptable alternative 3D reconstruction approach for curvilinear structures especially when acquiring large number of images is costly in terms of time and manual effort. Moreover, thanks to the recent advancements in deep neural networks, detection and matching tasks now may be combined in one neural network architecture and learned jointly. In this paper, we have therefore proposed a network architecture designed for matching curvilinear structures that delivers good performance in a wide variety of microscopy images while automatizing the process. We also introduced a novel loss term, \textit{variance loss}, which increased our network's localization capability. 
\section*{Acknowledgements}

Authors would like to thank Dr. Marc Legros for providing the HEA sample, Dr. Duncan T.L. Alexander for fruitful discussions, and Daniele Laub for help with TEM sample preparation. This work was supported in part by the Swiss National Science Foundation Sinergia grant Synergistic Approach to Capturing and Exploiting Microscopy Images.

\bibliography{bib/string,bib/biomed,bib/vision,bib/learning,bib/cime}

\begin{thebibliography}{10}
\expandafter\ifx\csname url\endcsname\relax
  \def\url#1{\texttt{#1}}\fi
\expandafter\ifx\csname urlprefix\endcsname\relax\def\urlprefix{URL }\fi
\expandafter\ifx\csname href\endcsname\relax
  \def\href#1#2{#2} \def\path#1{#1}\fi

\bibitem{Amelinckx74}
S.~Amelinckx, {The Characterization of Defects in Crystals}, Journal of Crystal
  Growth (1974) 6--10.

\bibitem{Hirsch2006a}
P.~Hirsch, D.~Cockayne, J.~Spence, M.~Whelan, {50 Years of {TEM} of
  Dislocations: Past, Present and Future}, Philosophical Magazine 86~(29-31)
  (2006) 4519--4528.

\bibitem{DeRosier1968}
D.~Rosier, D.~J., A.~Klug, {Reconstruction of Three Dimensional Structures from
  Electron Micrographs}, Nature 217~(5124) (1968) 130--134.

\bibitem{Weyland15}
M.~Weyland, P.~A. Midgley, {Chapter 6 Electron Tomography}, in:
  Nanocharacterisation (2), The Royal Society of Chemistry, 2015, pp. 211--299.

\bibitem{Midgley2009}
P.~A. Midgley, R.~E. Dunin-Borkowski, {Electron Tomography and Holography in
  Materials Science}, Nature Materials 8~(4) (2009) 271--280.

\bibitem{Sharp2008}
J.~H. Sharp, J.~S. Barnard, K.~Kaneko, K.~Higashida, P.~A. Midgley,
  {Dislocation Tomography Made Easy: {A} Reconstruction from {ADF-STEM} Images
  Obtained Using Automated Image Shift Correction}, in: Journal of Physics:
  Conference Series, 2008.

\bibitem{Barnard2006a}
J.~S. Barnard, J.~Sharp, J.~R. Tong, P.~A. Midgley, {High-Resolution
  Three-Dimensional Imaging of Dislocations}, Science 313~(5785) (2006) 319.

\bibitem{Jacome12}
A.~Jacome, G.~Eggeler, A.~Dlouhy, {Advanced Scanning Transmission Stereo
  Electron Microscopy of Structural and Functional Engineering Materials},
  Ultramicroscopy 122 (2012) 48--59.

\bibitem{Jacome18}
A.~Jacome, K.~Pöthkow, O.~Paetsch, H.-C. Hege, {Three-Dimensional
  Reconstruction and Quantification of Dislocation Substructures from
  Transmission Electron Microscopy Stereo Pairs}, Ultramicroscopy 195 (2018)
  157--170.

\bibitem{Oveisi17a}
E.~Oveisi, A.~Letouzey, D.~Alexander, Q.~Jeangros, R.~Schaublin, G.~Lucas,
  P.~Fua, C.~Hebert, {Tilt-Less {3D} Electron Imaging and Reconstruction of
  Complex Curvilinear Structures}, Nature Scientific Reports 7~(10630).

\bibitem{Oveisi18a}
E.~Oveisi, A.~Letouzey, S.~D. Zanet, G.~Lucas., M.~Antoni, P.~Fua, C.~Hebert,
  {Stereo-Vision Three-Dimensional Reconstruction of Curvilinear Structures
  Imaged with a {TEM}}, Ultramicroscopy 184~(A) (2018) 116--124.

\bibitem{Ronneberger15}
O.~Ronneberger, P.~Fischer, T.~Brox, U-net: Convolutional networks for
  biomedical image segmentation.

\bibitem{Hannah88}
M.~Hannah, {Digital Stereo Image Matching Techniques}, International Society
  for Photogrammetry and Remote Sensing 27~(III) (1988) 280--293.

\bibitem{Schonberger17}
J.~Sch{\"o}nberger, H.~Hardmeier, T.~Sattler, M.~Pollefeys, {Comparative
  Evaluation of Hand-Crafted and Learned Local Features}, in: Conference on
  Computer Vision and Pattern Recognition, 2017.

\bibitem{Medioni85}
G.~Medioni, R.~Nevatia, {Segment-Based Stereo Matching}, Computer Vision,
  Graphics, and Image Processing 31~(1) (1985) 2--18.

\bibitem{Ayache87}
N.~Ayache, F.~Lustman, {Fast and Reliable Passive Trinocular Stereovision}, in:
  International Conference on Computer Vision, 1987.

\bibitem{Meygret90}
A.~Meygret, M.~Thonnat, M.~Berthod, {A Pyramidal Stereovision Algorithm Based
  on Contour Chain Points}, in: European Conference on Computer Vision, 1990,
  pp. 83--88.

\bibitem{Fua93a}
P.~Fua, {A Parallel Stereo Algorithm That Produces Dense Depth Maps and
  Preserves Image Features}, Machine Vision and Applications 6~(1) (1993)
  35--49.

\bibitem{Boykov01b}
Y.~Boykov, O.~Veksler, R.~Zabih, {Fast Approximate Energy Minimization via
  Graph Cuts}, IEEE Transactions on Pattern Analysis and Machine Intelligence
  23~(11) (2001) 1222--1239.

\bibitem{Yang18d}
G.~Yang, H.~Zhao, J.~Shi, Z.~Deng, J.~Jia, {Segstereo: Exploiting Semantic
  Information for Disparity Estimation}, in: European Conference on Computer
  Vision, 2018, pp. 660--676.

\bibitem{Ji18}
M.~Ji, J.~Gall, H.~Zheng, Y.~Liu, L.~Fang, {Surfacenet: An End-To-End 3D Neural
  Network for Multiview Stereopsis}, in: International Conference on Computer
  Vision, 2017, pp. 2326--2334.

\bibitem{Huang18b}
P.~Huang, K.~Matzen, J.~Kopf, N.~Ahuja, J.~Huang, {Deepmvs: Learning Multi-View
  Stereopsis}, in: Conference on Computer Vision and Pattern Recognition, 2018,
  pp. 2821--2830.

\bibitem{Kar17}
A.~Kar, C.~H{\"a}ne, J.~Malik, {Learning a Multi-View Stereo Machine}, in:
  Advances in Neural Information Processing Systems, 2017, pp. 364--375.

\bibitem{LeCun15}
Y.~LeCun, Y.~Bengio, G.~Hinton, {Deep Learning}, Nature 521 (2015) 436--444.

\bibitem{Bromley93}
J.~Bromley, I.~Guyon, Y.~LeCun, E.~S\"{a}ckinger, R.~Shah, {Signature
  verification using a "Siamese" time delay neural network} (1993) 737–744.

\bibitem{Kendall17a}
A.~Kendall, H.~Martirosyan, S.~Dasgupta, P.~Henry, {End-To-End Learning of
  Geometry and Context for Deep Stereo Regression}, in: International
  Conference on Computer Vision, 2017.

\bibitem{Lowe04}
D.~G. Lowe, {Distinctive Image Features from Scale-Invariant Keypoints},
  International Journal of Computer Vision 20~(2) (2004) 91--110.

\bibitem{Bolles86}
R.~Bolles, R.~Horaud, {3DPO: A Three-Dimensional Part Orientation System},
  International Journal of Robotics Research 5~(3) (1986) 3--26.

\bibitem{Liang18}
Z.~Liang, Y.~Feng, Y.~Guo, H.~Liu, W.~Chen, L.~Qiao, L.~Zhou, J.~Zhang,
  {Learning for Disparity Estimation through Feature Constancy}, in: Conference
  on Computer Vision and Pattern Recognition, 2018.

\bibitem{Pang17}
J.~Pang, W.~Sun, J.~S. Ren, C.~Yang, Q.~Yan, {Cascade Residual Learning: A
  Two-stage Convolutional Network for Stereo Matching}, in: International
  Conference on Computer Vision Workshops, 2017.

\bibitem{KingmaB14}
D.~P. Kingma, J.~Ba, Adam: {A} method for stochastic optimization, in:
  Y.~Bengio, Y.~LeCun (Eds.), 3rd International Conference on Learning
  Representations, {ICLR} 2015, San Diego, CA, USA, May 7-9, 2015, Conference
  Track Proceedings, 2015.

\bibitem{OVEISI2019139}
E.~Oveisi, M.~C. Spadaro, E.~Rotunno, V.~Grillo, C.~Hebert, {Insights into
  Image Contrast from Dislocations in {ADF-STEM}}, Ultramicroscopy 200 (2019)
  139--148.

\end{thebibliography}
\newpage
\section*{Appendix} 

\subsection*{Neural Network Training}

All images are resized to 512x512 and normalized to (0, 1). In order to increase dataset size, we have applied 2D image augmentations: random scaling, rotation, brightness with parameters (0.7, 1.2), (-60, 60), (0.8, 1.2) respectively. All models are optimized using the  Adam algorithm~\cite{KingmaB14}  with beta parameters=(0.9, 0.999) and learning rates 0.001 and 0.0001 for detection and stereo matching networks respectively. Training was performed on Tesla V100. We used the PyTorch neural network framework for our implementation.

\begin{figure}[H]
	\begin{tabular}{cc}
		\includegraphics[width=40mm]{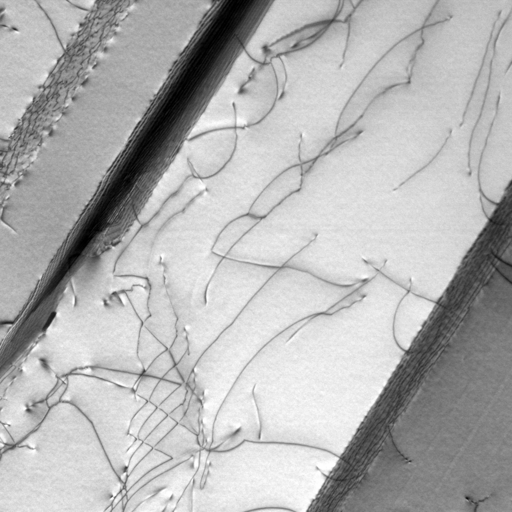} &   \includegraphics[width=40mm]{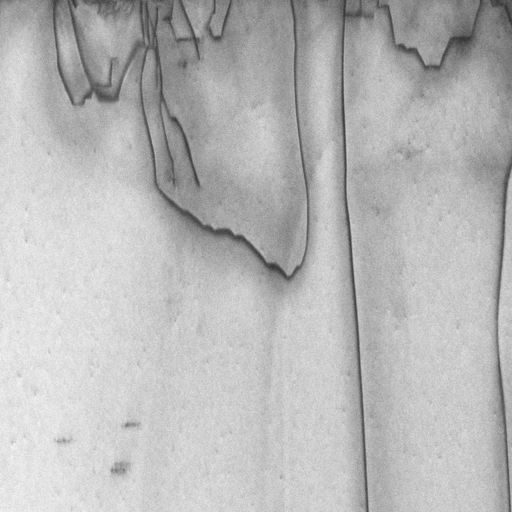} \\
		\includegraphics[width=40mm]{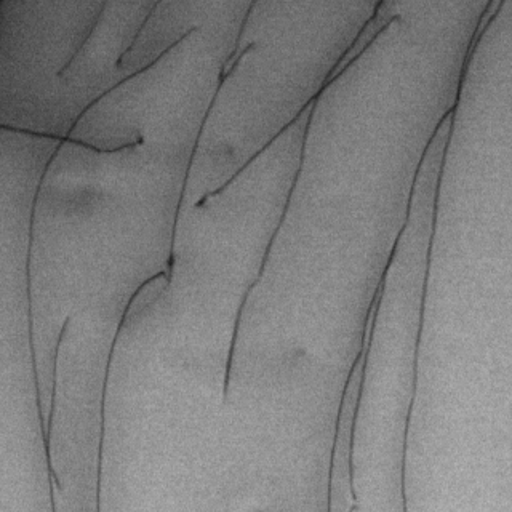} &   \includegraphics[width=40mm]{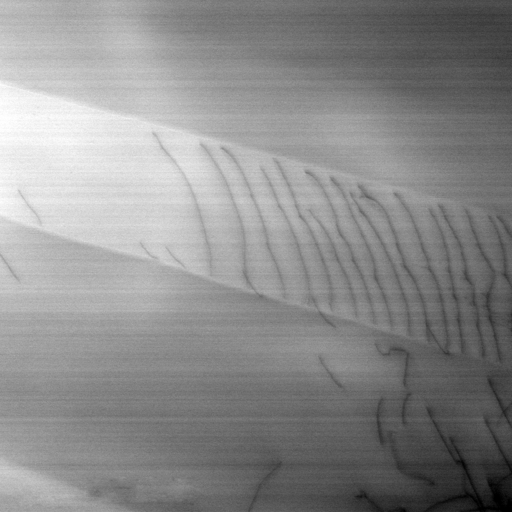} \\
	\end{tabular}
	\caption{Samples from our dataset.}
	\label{samples}
\end{figure}

\subsection*{Sample Preparation and Imaging Conditions}

\begin{itemize}

    \item {\bf TiAl sample.}
    
Dataset of dislocations in TiAl specimen was used for training and validation, as well as to test the 3D reconstruction performed by 3D CNN.
    
Dislocations were imaged in a TiAl super-alloy thin sample with nominal bulk composition of Ti-46.8Al-1.7Cr-1.8Nb (at.$\%$). To prepare electron transparent foils from the bulk sample, the specimen was first mechanically thinned with diamond discs and electropolished (electrolyte bath composition: 5 vol.$\%$ perchloric acid, 35 vol.$\%$ 1-butanol and 60 vol.$\%$ methanol; voltage: 35 V). Then ion milling with high energy gallium ions was used to further decrease the thickness.
Dislocation imaging was performed on a Thermo Scientific Tecnai Osiris transmission electron microscope operated in scanning mode (STEM) at 200 kV. Images were acquired in annular bright-field (ABF) configuration for which a 70 $\mu$m condenser aperture was used to form the probe with 12.4 mrad convergence semi-angle and detector collection angle was set to 18.4 mrad so that the rim of the direct disc (000) covers the annular detector.  Tilt-series of ABF-STEM images with 1024x1024 pixels resolutions was acquired while the sample was tilted within $\pm$50 degree tilt range and imaged every 2 degree with slightly positive deviation from the $\textbf{g}$=(002) two-beam diffraction condition. The sample was carefully oriented using a Fischione Instruments dual-axis tomography holder (model 2040) so that the deviation parameter $\textit{s}_\textbf{g}$ remains constant throughout the tilt range.

\begin{figure}[H]
	\centering
	\begin{tabular}{cc}
		\includegraphics[width=80mm]{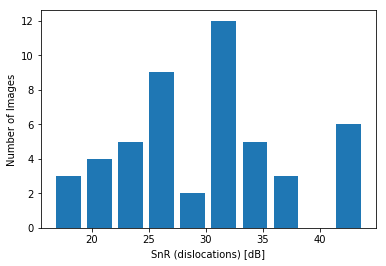} \\   \includegraphics[width=80mm]{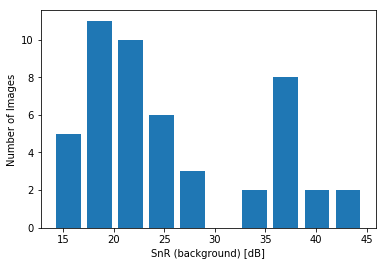} \\
		\includegraphics[width=80mm]{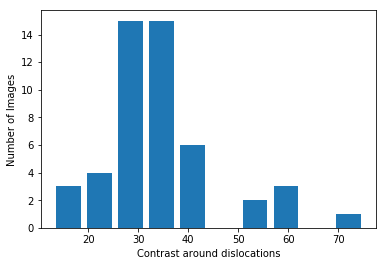}   
	\end{tabular}
	\caption{Training dataset distribution with respect to signal-to-noise ratio and contrast measurements.}
	\label{stats}
\end{figure}   
    \item {\bf GaN sample.}
    
Dataset of dislocation images used for training and validation was acquired from a GaN sample. Details on sample preparation and imaging condition can be found in \cite{Oveisi17a,OVEISI2019139}.
    
    \item {\bf Cantor alloy sample.}
    
Images of dislocations from high-entropy Cantor alloy (equimolar CrCoFeMnNi) used for training and validation. Sample has been homogenized 2 hours at 1000$^{\circ}$C under vacuum and then 1x3 mm rectangles mechanically thinned down to 30-50 $\mu$m with SiC papers and electropolished (electrolyte bath composition: 90$\%$ ethanol and 10$\%$ perchloric acid).
Dislocation imaging in Cantor alloy was performed on a Thermo Scientific Themis Titan transmission electron microscope in bright-field (BF) configuration. 

\end{itemize}

\subsection*{Dataset Specification}
To be able to train models that generalize well on broad range of TEM images of disloations, we constructed a dataset that is diverse both visually and geometrically. Our dislocation dataset consists of images of different material samples with varying imaging conditions. In Fig. \ref{samples} we present 4 data samples to show these variations in image characteristics in the dataset. Moreover, in Fig \ref{stats} distribution of images with respect to their signal-to-noise ratio  and contrast measurements are shown. We measure SnR values  separately for dislocation and background regions on an image. Contrast is measured as average differene in grayscale values between dislocation cores and background pixels at vicinity of dislocations. In Table \ref{tb:snr_test}, we show these measurements also for test samples we report our results. \\
\begin{table}[ht]
	\centering
	\resizebox{90mm}{!}{
\begin{tabular}{|l|c|c|c|}
	\toprule
	{Sample/Metric} &  { SnR (disloc.) [dB]} & { SnR (backgr.) [dB] } & {Contrast}\\
	\hline
	GaN & 30.16 & 44.68 & 42.64 \\
	\hline
	TiAl & 34.24 & 25.32 & 35.78  \\
	\hline
	Cantor Alloy & 25.70 & 16.42 & 26.91 \\
	\bottomrule
\end{tabular}
}
\caption{ Signal-to-noise ratio and contrast values of the test samples that we report our results.} 
\label{tb:snr_test}
\end{table}

\begin{figure*}[t]
\centering

\includegraphics[width=1\hsize]{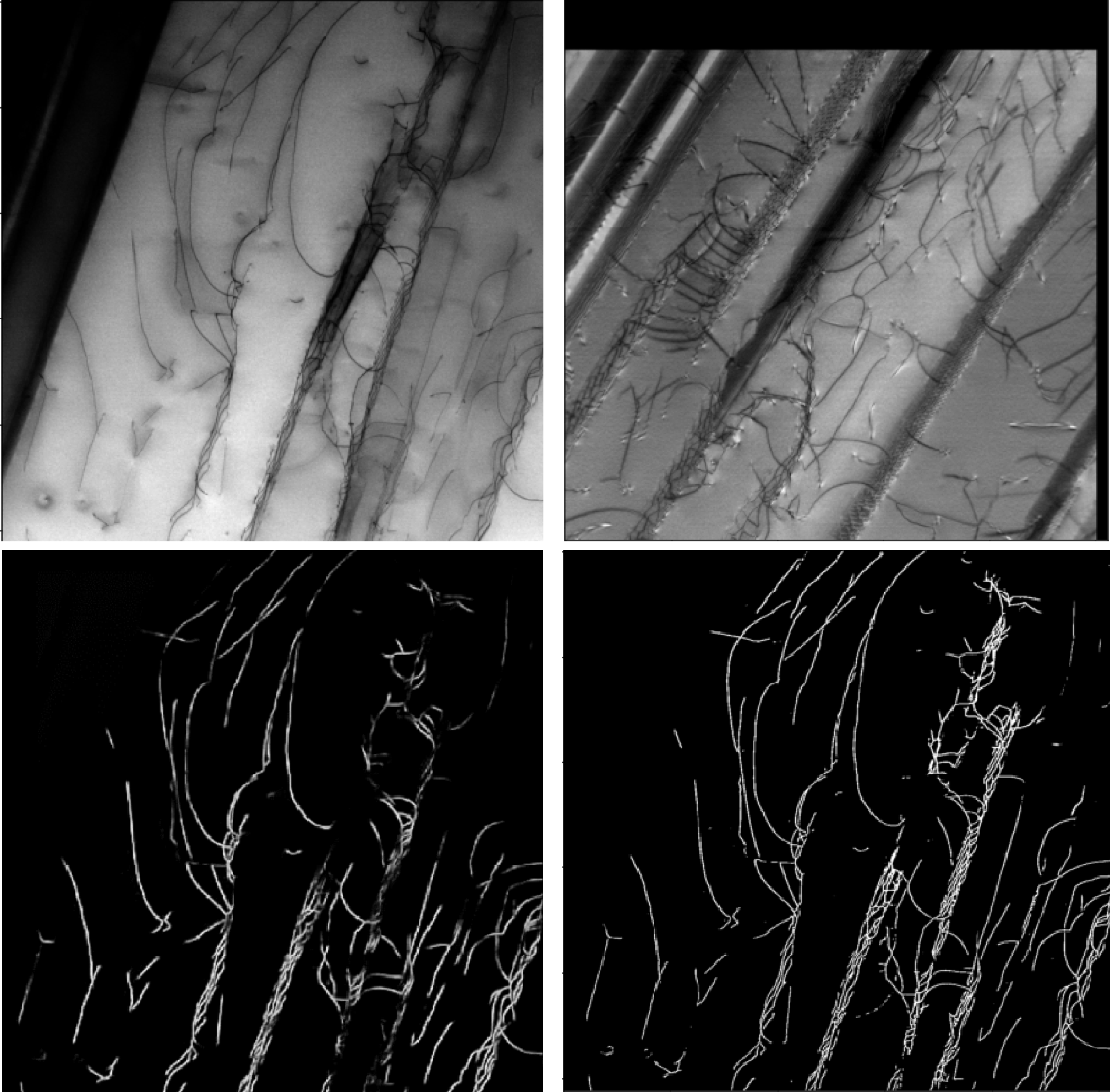}

\caption{We show contour detection network's (UNet) output for TiAl test sample (top left) before (bottom left) and after (bottom right) adding additional low-contrast training sample (top right) to the training set. The performance significantly improved after adding only 1 image obtained from a different image sequence. It could be seen that the network’s output was generally worse on low-contrast areas of the image. Adding one image having similar visual conditions to the training set eliminates detection errors on these regions and improves overall performance.}
\label{fig:appendix_imp}
\end{figure*}
\end{document}